\documentclass[12pt,prb,aps]{revtex4-1}
\usepackage{graphicx}
\usepackage{rotating}
\usepackage{array}
\usepackage{amsmath}
\usepackage{multirow}
\usepackage{setspace}
\usepackage{soul}
\usepackage{color}
\usepackage{xcolor}

\begin{document}
\title{Non-iterative triples excitations in equation-of-motion coupled-cluster theory 
for electron attachment with applications to bound and temporary anions}
% treatment of 

\author{Thomas-C. Jagau}
\affiliation{\small Department of Chemistry, University of Munich (LMU), D-81377 Munich, Germany}

\begin{abstract}
The impact of residual electron correlation beyond the equation-of-motion coupled-cluster singles 
and doubles approximation (EOM-CCSD) on positions and widths of electronic resonances 
is investigated. To establish a method that accomplishes this task in an economical manner, 
several approaches proposed for the approximate treatment of triples excitations are reviewed 
with respect to their performance in the electron attachment (EA) variant of EOM-CC theory. 

The recently introduced EOM-CCSD(T)(a)* method, which includes non-iterative corrections 
to the reference and the target states, reliably reproduces vertical attachment energies from 
EOM-EA-CC calculations with singles, doubles, and full triples excitations in contrast to schemes
in which non-iterative corrections are applied only to the target states. 

Applications of EOM-EA-CCSD(T)(a)* augmented by a complex absorbing potential (CAP) to 
several temporary anions illustrate that shape resonances are well described by EOM-EA-CCSD, 
but that residual electron correlation often makes a non-negligible impact on their positions and 
widths. The positions of Feshbach resonances, on the other hand, are significantly improved 
when going from CAP-EOM-EA-CCSD to CAP-EOM-EA-CCSD(T)(a)*, but the correct energetic 
order of the relevant electronic states is still not achieved.

% comparable in size to the perturbation by the CAP
% Several approaches for the approximate treatment of triples excitations are applied to the 
% equation-of-motion coupled-cluster method for electron attachment. 

% "benchmark"
% and the EOM-EA-CC3 method, which features iterative corrections, 
 
\end{abstract}
\maketitle

\clearpage

\section{Introduction} 
\label{sec:int}
Quantum chemistry of bound electronic states has reached a level where quantitative agreement 
between theory and experiment is possible for many observables.\cite{elstbook} The same has so 
far not been achieved for processes involving electronic resonances,\cite{jagau17} i.e., autoionizing 
states with finite lifetime embedded in the continuum. Although theoretical methods for resonances 
possess considerable predictive power,\cite{jagau17} sizable discrepancies between theory and 
experiment often still exist. Sources of uncertainty on the side of theory are manifold and can 
include errors introduced by improper handling of the resonance character, neglect of structural 
relaxation and vibrational effects, insufficient basis-set convergence, and incomplete treatment 
of electron correlation. 
% for example, for thermochemical parameters or rotational spectroscopy

The latter issue motivates this article. While the general importance of electron correlation 
for resonances is well established,\cite{jagau17,sommerfeld98,white17} a quantification of 
higher-order correlation effects as routinely possible for bound states, has never been done for 
resonances. This is because their theoretical treatment is more demanding than that of bound 
states. Electronic resonances do not represent discrete $L^2$ integrable states in Hermitian 
quantum mechanics and only in a non-Hermitian formalism is it possible to treat them as discrete 
states with complex energy.\cite{nhqmbook,jagau17} Strategies to compute complex resonance 
energies with bound-state electronic-structure methods include stabilization methods,\cite{hazi70,
nestmann85}, analytic continuation of the coupling constant,\cite{horacek10} and especially 
complex-variable techniques such as complex scaling,\cite{aguilar71,balslev71,simon79,
moiseyev79}, complex basis functions, \cite{mccurdy78} and complex absorbing potentials 
(CAPs).\cite{jolicard85,riss93} Particular advantages of the latter approach are its compatibility 
with the Born-Oppenheimer approximation and the possibility to extend techniques for computing 
molecular properties \cite{jagau16} and analytic energy gradients \cite{benda17} to electronic 
resonances.

%For bound states, electron correlation effects can be quantified 
%to very high precision for energies and properties, for example, using the coupled-cluster (CC) 
%hierarchy of methods, while the impact of higher-order correlation on resonances is less well 
%even though the general importance 
% whereas for resonances, the 
%impact beyond the EOM-CC singles and doubles approximation (EOM-CCSD) is unclear. 
%even though there are some 
%investigations based on multireference configuration interaction theory. 

%While complex-variable techniques have been combined with various electronic-structure 
%methods,\cite{} realizations within 
When aiming at a high-accuracy treatment of electronic resonances in polyatomic molecules, 
the combination of complex-variable techniques with equation-of-motion coupled-cluster 
(EOM-CC) theory offers several formal advantages and is also numerically promis\-ing.\cite{
ghosh12,bravaya13,zuev14,jagau16b,white17} The EOM-CC formalism \cite{emrich80,geertsen89,
stanton93,ccbook,krylov08,sneskov12} allows to treat electronically excited \cite{stanton93} 
(EOM-EE), ionized \cite{stanton94} (EOM-IP), and electron-attached \cite{nooijen95} (EOM-EA) 
states as eigenfunctions of the same effective Hamiltonian, which is crucial for electronic 
resonances because it enables a consistent description of a resonance relative to the 
ionization and detachment continua.\cite{jagau14} Furthermore, EOM-CC treats states 
with different character on an equal footing and by including higher excitations in the 
ansatz for the wave function, the description can be systematically improved up to the 
full configuration interaction limit. 
% on a starting from the same CC reference state. extends the CC hierarchy of methods to multistate problems and 
%Moreover, states with multi-configurational 

Whereas electronic resonances so far have not been studied beyond the EOM-CC singles 
and doubles (EOM-CCSD) approximation, the use of more accurate schemes is commonplace 
for bound states. EOM-CCSD enables a reliable treatment of many systems --provided that 
the reference state has single-reference character-- but it has become clear that higher 
excitations significantly impact energies and molecular properties and cannot be neglected 
if quantitative accuracy is sought. Besides the EOM-CCSDT family of methods,\cite{kucharski01,
musial03a,musial03b} general CC implementations that allow to consider arbitrarily high 
orders in the EOM-CC hierarchy \cite{hirata00a,hirata00b,hirata04,kallay04} are especially 
noteworthy. Since already the CCSDT method scales as $\mathcal{O}(N^8)$ with $N$ as 
the size of the basis set (compared to $\mathcal{O}(N^6)$ for CCSD), numerous approaches 
for the approximate consideration of triples excitations have been proposed. In particular, 
EOM-CC3 \cite{christiansen95,koch97} has been established as an intermediate method 
between EOM-CCSD and EOM-CCSDT entailing iterative $\mathcal{O}(N^7)$ cost. It should 
be also mentioned in this context that an implementation of a particular EOM-EE-CC method 
provides access to the corresponding EOM-IP and EOM-EA states by means of the 
continuum-orbital trick \cite{stanton99} albeit such an approach entails higher than necessary 
computational cost.  
% watts94,kowalski01, EOM-CC(2,3)

There have also been many efforts \cite{watts95,watts96,christiansen96,stanton96,saeh99,
hirata01,kowalski04,manohar08,manohar09,matthews16} to devise EOM-CC methods 
that account for the effect of triples excitations in a non-iterative fashion similar to CCSD(T) 
\cite{raghavachari89,bartlett90} for ground-state energies and properties. It has become
apparent, however, that a balanced correction of several states is more difficult to achieve 
than that of a single state. Consequently, non-iterative triples corrections clearly improve 
the description of the target states, whereas results for energy differences are mixed. Also, 
while there have been several benchmark studies on excitation energies,\cite{sauer09,
watson13,kannar17} and also systematic investigations of ionized states \cite{saeh99,
manohar09,matthews16}, none of the non-iterative triples corrections has been used 
within the EA variant of EOM-CC. 

EOM-EA-CC is well suited for radical anions and such species often have resonance 
character.\cite{simons08} Since all complex-variable techniques entail considerably 
increased computational cost compared to the corresponding bound-state methods,\cite{
jagau17} non-iterative triples corrections to EOM-EA-CCSD appear as an attractive approach 
for studying the impact of higher excitations on electronic resonances. 

The purpose of this article is hence twofold: Establishing a method for the economical treatment 
of triples excitations in EOM-EA-CC and applying that method to investigate the impact of 
residual electron correlation beyond EOM-EA-CCSD on resonance positions and widths. 
As for the first objective, EOM-CCSD* \cite{stanton96,saeh99} and EOM-CCSD(fT) \cite{
manohar08,manohar09} are investigated as examples of methods that correct only the target 
state and the novel EOM-CCSD(T)(a)* method \cite{matthews16} is considered because 
it includes a correction to the CCSD reference state as well. As for the second objective, 
different effects can be anticipated for shape and Feshbach resonances. Radical anions 
of the former type are of single-attachment character and should be well described by 
EOM-EA-CCSD, whereas those of the second type are dominated by double excitations 
and their description should be improved significantly when going beyond EOM-EA-CCSD. 

The remainder of the article is structured as follows: Section \ref{sec:th} presents a brief revision 
of the relevant parts of EOM-CC theory and the non-iterative triples corrections investigated 
here with a special focus on their EA variants and the application to resonance states using 
CAPs. In Section \ref{sec:app1}, the performance of the various approximate methods is 
assessed via application to a test set of bound electron-attached states and comparison to full 
EOM-EA-CCSDT. Sections \ref{sec:app2} and \ref{sec:app3} present a number of illustrative 
applications to shape and Feshbach resonances, respectively, while Section \ref{sec:con} 
gives some concluding remarks. 

\section{Theory} 
\label{sec:th}
In EOM-CC theory, \cite{ccbook,krylov08,sneskov12} the wave function of a target state $|\Psi \rangle$ 
is parametrized as 
\begin{align} \label{sgr}
| \Psi \rangle &= R \, e^T | \Phi_0 \rangle \\
\langle \Psi | &=  \langle \Phi_0 | L \, e^{-T} 
\label{sgl} \end{align}
with $e^T | \Phi_0 \rangle$ as the CC reference state and $| \Phi_0 \rangle$ as a single Slater 
determinant that usually fulfills the Hartree-Fock (HF) equations. The cluster operator $T$ is 
defined as 
\begin{equation}
T = T_1 + T_2 + T_3 + \dots = \sum_{ai} t_i^a \, a^\dagger_a a_i + \frac{1}{4} 
\sum_{abij} t_{ij}^{ab} \, a^\dagger_a a_i a^\dagger_b a_j + \frac{1}{36} \sum_{abcijk} 
t_{ijk}^{abc} \, a^\dagger_a a_i a^\dagger_b a_j a^\dagger_c a_k + \dots
\end{equation}
where the indices $a, b, c, \dots$ and $i, j, k, \dots$ denote virtual and occupied spin orbitals and 
$a^\dagger$ and $a$ represent second-quantized creation and annihilation operators. $T$ is 
determined from the CC equations 
\begin{align} \label{eqcc1}
\langle \Phi_{ijk\dots}^{abc\dots} | e^{-T} H e^T | \Phi_0 \rangle &= 0, \\
\langle \Phi_0 | e^{-T} H e^T | \Phi_0 \rangle &= E_\text{ref} \label{eqcc2}
\end{align}
with $E_\text{ref}$ as the energy of the CC reference state.

The excitation operator $R$ and its de-excitation counterpart $L$ are chosen in different ways 
depending on the desired target states. In this article, the focus is on the EA variant of EOM-CC, 
where $R$ and $L$ are chosen as 
\begin{align}
R &= R_1 + R_2 + R_3 + \dots = \sum_{a} r^a a^\dagger_a 
+ \frac{1}{2} \sum_{abi} r_{i}^{ab} a^\dagger_a a_i a^\dagger_b + \frac{1}{12} \sum_{abcij} r_{ij}^{abc} 
a^\dagger_a a_i a^\dagger_b a_j a^\dagger_c + \dots~, \\
L &= L_1 + L_2 + L_3 + \dots = \sum_{a} l_a a_a 
+ \frac{1}{2} \sum_{abi} l_{ab}^{i} a_i^\dagger a_a a_b + \frac{1}{12} \sum_{abcij} l_{abc}^{ij} 
a_i^\dagger a_a a_j^\dagger a_b a_c + \dots
\end{align}
i.e., they describe the addition of an electron to the system and yield attachment energies. 
Other choices of $R$ and $L$ provide access to electronic excitations, ionization, double 
ionization, double attachment as well as spin-flipping manifolds.\cite{krylov08}

%In EOM-EE-CC, $R$ is a particle-conserving operator 
%\begin{align}
%R^\text{EE} &= R_0 + R_1 + R_2 + R_3 + \dots \nonumber \\
%& = r_0 + \sum_{ai} r_i^a a^\dagger_a a_i + \frac{1}{4} \sum_{abij} r_{ij}^{ab} a^\dagger_a a_i 
%a^\dagger_b a_j + \frac{1}{36} \sum_{abcijk} r_{ijk}^{abc} a^\dagger_a a_i a^\dagger_b a_j 
%a^\dagger_c a_k + \dots
%\end{align}
%that provides access to excitation energies. In EOM-EA-CC, $R$ is chosen as 
%\begin{equation}
% EOM-IP-CC and 
%R^\text{IP} &= R_1^\text{IP} + R_2^\text{IP} + R_3^\text{IP} + \dots = \! \sum_i r_i a_i + \frac{1}{2} 
%\sum_{aij} r_{ij}^{a} a^\dagger_a a_i a_j + \frac{1}{12} \sum_{abijk} r_{ijk}^{ab} a^\dagger_a a_i 
%a^\dagger_b a_j a_k + \dots \\
%R^\text{EA} \!= \!R_1^\text{EA} \!+ \!R_2^\text{EA} \!+ \!R_3^\text{EA} \!+ \dots = \!\sum_{a} r^a a^\dagger_a 
%+ \frac{1}{2} \sum_{abi} r_{i}^{ab} a^\dagger_a a_i a^\dagger_b + \frac{1}{12} \sum_{abcij} r_{ij}^{abc} 
%a^\dagger_a a_i a^\dagger_b a_j a^\dagger_c + \dots
%\end{equation}
%i.e., it adds one electron to the system, respectively, and thus allows to determine 
%ionization and attachment energies. Further choices of $R$ describe double ionization and double 
%attachment as well as spin-flipping processes.

To determine the energy of the target state and the amplitudes in the operator $R$, Eq. \eqref{sgr} 
is plugged into the Schr\"odinger equation, premultiplied by $e^{-T}$ and projected onto the set of 
determinants $\langle \Phi_{ij\dots}^{abc\dots}|$. This yields the eigenvalue equation 
\begin{align} \label{eqeomr}
\langle \Phi^{a} | \overline{H}\, R | \Phi_0 \rangle &= \omega \, r^{a} ~ \forall \; \langle \Phi^a | ~,\\
\langle \Phi_{i}^{ab} | \overline{H}\, R | \Phi_0 \rangle &= \omega \, r_{i}^{ab} ~ \forall \; 
\langle \Phi_i^{ab} | ~,\nonumber \\
\dots \nonumber
\end{align}
for the similarity-transformed Hamiltonian $\overline{H} = e^{-T} H e^T - E_\text{ref}$ with $\omega = 
E - E_\text{ref}$.
%with $\omega$ as excitation energy (or ionization or attachment energy in EOM-IP-CC and EOM-EA-CC). 
The corresponding eigenvalue equation for the left-hand side reads 
\begin{align}  \label{eqeoml}
\langle \Phi_0 | L\, \overline{H} | \Phi^a \rangle &= \omega \, l_a~ \forall \; |\Phi^a \rangle ~, \\
\langle \Phi_0 | L\, \overline{H} | \Phi_i^{ab} \rangle &= \omega \, l_{ab}^i~ \forall \; |\Phi_i^{ab} \rangle~, \nonumber \\
\dots \nonumber
\end{align}
Since $\overline{H}$ is not Hermitian, its right and left eigenstates are not complex conjugates of each 
other, but they can be chosen to be biorthonormal.  

Truncating $T=T_1+T_2$, $R=R_1+R_2$, $L=L_1+L_2$ together with corresponding truncations 
of the projection manifolds defines the EOM-CCSD model. As outlined in Section \ref{sec:int}, 
different strategies can be pursued to take account of triples excitations in a non-iterative manner.
For the approaches that are investigated here, i.e., EOM-CCSD*, EOM-CCSD(fT), and EOM-CCSD(T)(a)*, 
working equations have been derived previously in the context of EOM-EE-CC and EOM-IP-CC and 
the reader is referred to the original articles \cite{stanton96,saeh99,manohar08,manohar09,matthews16} 
for a comprehensive discussion. 

The adaptation to EOM-EA-CC is straightforward; in particular, the EOM-EA-CCSD* and 
EOM-EA-CCSD(fT) energy corrections take on the form 
\begin{equation} \label{eqdelta}
\Delta E =  \frac{1}{12} \, \sum_{abcij} \, \frac{l_{abc}^{ij} \cdot r_{ij}^{abc}}{\varepsilon_i + \varepsilon_j 
- \varepsilon_a - \varepsilon_b - \varepsilon_c + \omega}
\end{equation}
with $\varepsilon_i, \varepsilon_j, \dots$ as orbital energies. Explicit expressions for the approximate 
triples amplitudes $r_{ij}^{abc}$ and $l_{abc}^{ij}$ are obtained for EOM-EA-CCSD(fT) as
\begin{align} \label{eqft1}
r^{abc}_{ij} &= \langle \Phi_{ij}^{abc} |\, \overline{H} \,(R_1 + R_2) | \Phi_0 \rangle~, \\ \label{eqft2}
l_{abc}^{ij} &= \langle \Phi_0 | (L_1 + L_2) \, \overline{H} \, | \Phi_{ij}^{abc} \rangle~. 
\end{align}
In EOM-EA-CCSD*, $r^{abc}_{ij}$ and $l_{abc}^{ij}$ are further approximated by considering only 
those terms in Eqs. \eqref{eqft1} and \eqref{eqft2} that contribute to $\Delta E$ in third order of the 
correlation perturbation.\cite{stanton96,saeh99} Evaluation of Eq. \eqref{eqdelta} scales as 
$\mathcal{O}(N^6)$ for both approaches and is thus not a rate limiting step since the determination 
of the CCSD reference state already entails $\mathcal{O}(N^6)$ cost and Eqs. \eqref{eqeomr} and 
\eqref{eqeoml} scale as $\mathcal{O}(N^5)$ in EOM-EA-CCSD. 
%\begin{align}  \label{eqstar}
%r^{abc}_{ij} &= \sum_e l_{ce}^j \\
%l_{abc}{ij} &= \nonumber
%\end{align}

The EOM-EA-CCSD(T)(a)* approach is more elaborate than EOM-EA-CCSD* and EOM-EA-CCSD(fT) 
in that it involves a correction to the cluster operator $T$ from which $\overline{H}$ is constructed. 
\cite{matthews16} Specifically, an approximate $T_3$ is obtained from a lowest-order triples amplitude 
equation 
\begin{equation} \label{eqtamp}
0 = \langle \Phi_{ijk}^{abc} | [W, T_2] + [F, T_3^{[2]}] | \Phi_0 \rangle
\end{equation}
with $F$ and $W$ as one-particle and two-particle parts of the Hamiltonian. 
%the CCSDT-1 \cite{lee84}
%The amplitudes of $T_3^{[2]}$ become
%\begin{equation} \label{eqta3}
%T_3^{[2]} = \frac{1}{36} \sum_{abcijk} \sum_{abcijk}
%t_{ijk}^{abc[2]} =  \frac{P(k/ij) P(a/bc) \Big( \sum_e \langle bc || ek \rangle \, t_{ij}^{ae} \Big) 
%- P(i/jk) P(c/ab) \Big( \sum_m \langle mc || jk \rangle \, t_{im}^{ab} \Big)}{\varepsilon_i + \varepsilon_j 
%+ \varepsilon_k - \varepsilon_a - \varepsilon_b - \varepsilon_c}
%\end{equation}
%with $P(p/qr)X(pqr) = X(pqr) - X(qpr) - X(rqp)$ for a generic function $X$. 
These approximate triples amplitudes are then used to correct the converged CCSD $T_1$ 
and $T_2$ amplitudes for the effect of $T_3$ according to 
\begin{align} \label{eqta1}
%= t_i^a + \frac{1}{4} \cdot \frac{\sum_{mnef} \langle mn || ef \rangle \, t_{imn}^{aef}}
%{\varepsilon_i - \varepsilon_a}
t_i^{a\, \text{corr}} &= t_i^a+ \frac{\langle \Phi_i^a | [W, T_3^{[2]}] |\Phi_0 \rangle}{\varepsilon_i - 
\varepsilon_a} ~,\\ \label{eqta2}
t_{ij}^{ab\,\text{corr}} &= t_{ij}^{ab} + \frac{\langle \Phi_{ij}^{ab} | [W, T_3^{[2]}] |\Phi_0 \rangle}
{\varepsilon_i + \varepsilon_j - \varepsilon_a - \varepsilon_b}~. %\\
%&= t_{ij}^{ab} + \frac{1}{2} \cdot \frac{P(ab) \sum_{mef} \langle bm || ef \rangle \, t_{iim}^{aef[2]} 
%- P(ij) \sum_{mne} \langle mn || je \rangle \, t_{imn}^{abe[2]}}{\varepsilon_i + \varepsilon_j 
%- \varepsilon_a - \varepsilon_b} \nonumber
\end{align}
%with $P(pq)X(pq)= X(pq) - X(qp)$. 
$E_\text{ref}$ is evaluated from Eq. \eqref{eqcc2} using $T_1^\text{corr}$ and $T_2^\text{corr}$. 
The EOM amplitudes are determined via Eqs. \eqref{eqeomr} and \eqref{eqeoml} from 
$\overline{H}^\text{corr} = e^{-T_1^\text{corr} - T_2^\text{corr}} (H - E_\text{ref}) \, e^{T_1^\text{corr} 
+ T_2^\text{corr}}$ including an additional direct contribution of $T_3^{[2]}$, which in EOM-EA-CC 
takes on the form  
\begin{align} \label{eqtah1}
\langle \Phi_i^{ab} | \overline{H} R_1 |\Phi_0 \rangle &\Leftarrow 
\langle \Phi_i^{ab} | [W, T_3^{[2]}] R_1 |\Phi_0 \rangle~, \\  \label{eqtah2}
% %= \sum_e r^e \langle || \rangle t_{}^{[2]}
\langle \Phi_0 | L_2 \overline{H} |\Phi^a \rangle &\Leftarrow 
\langle \Phi_0 | L_2 [W, T_3^{[2]}] | \Phi^a \rangle~.
%= \frac{1}{2} \sum_{efm} l_{ef}^m \langle || \rangle t_{}^{[2]}
\end{align}
The final EOM-EA-CCSD(T)(a)* energy is obtained by invoking Eq. \eqref{eqdelta} in analogy 
to EOM-EA-CCSD*.\cite{matthews16} However, in contrast to the latter method, the overall 
scaling of EOM-EA-CCSD(T)(a)* is determined by Eq. \eqref{eqtamp} and thus non-iterative 
$\mathcal{O}(N^7)$. 

In EOM-EA-CC, all of the non-iterative triples corrections are size-intensive because the target 
states are decoupled from the reference state. Also, the energy correction obtained through Eq. 
\eqref{eqdelta} is invariant to rotations among occupied and virtual orbitals for all methods. It 
should be added, however, that canonical HF orbitals were assumed in all equations and 
additional terms may appear if other orbitals are used. 

% eqdelta entails that target energies and ref energies are not eigenfunctions to the same H, resp+prop difficult
% advantage of EOM-CC3: properties via response + derivative theory

%This yields the 
%following explicit expressions in terms of spin orbitals 
%\begin{align}
%\Delta E^* &=  \frac{1}{12} \sum_{abcij} \frac{}{\epsilon_a + \epsilon_b + \epsilon_c - \epsilon_i - \epsilon_j + \omega} \\
%\Delta E^{fT} &=  \frac{1}{12} \sum_{abcij} \frac{\langle \Phi_0 | (L^\text{EA}_1 + L^\text{EA}_2) \bar{H} | 
%\Phi_{ij}^{abc} \rangle \cdot \langle \Phi_{ij}^{abc} | \bar{H} (R^\text{EA}_1 + 
%R^\text{EA}_2)| \Phi_0 \rangle}{\epsilon_a + \epsilon_b + \epsilon_c - \epsilon_i - \epsilon_j + \omega} 
%&= \frac{1}{12} \sum_{abcij} {}{\epsilon_a + \epsilon_b + \epsilon_c - \epsilon_i - \epsilon_j + \omega} 
%\end{align}

To extend the scope of EOM-CC theory to electronic resonances, a CAP is added to the molecular 
Hamiltonian \cite{jolicard85,riss93}: 
\begin{equation} \label{eq:cap1}
H(\eta) = H_0 - i \, \eta \, W~.
\end{equation}
$H(\eta)$ has complex eigenvalues $E - i \, \Gamma/2$, from which the resonance positions $E$ and 
resonance widths $\Gamma$ are obtained.\cite{jagau17} Although the form of the working equations 
need not be modified if the CAP is added at the HF stage as it is the case in the present implementation, 
all wave function parameters become complex-valued and, moreover, one needs to replace the usual 
scalar product by the c-product in all equations because $H(\eta)$ is not Hermitian but complex-symmetric. 
\cite{moiseyev78} $W$ is chosen as a shifted quadratic potential here and the optimal value of the CAP 
strength $\eta$ is determined by a perturbative analysis of $E(\eta)$.\cite{riss93} This entails recalculating 
the energy for many values of $\eta$ and is the main reason for the increased computational cost of CAP 
methods. The lowest order of perturbation theory yields the criterion $\text{min}[\eta \, dE/d\eta]$, but 
superior results are obtained when removing this term from the energy and considering the corrected 
energy \cite{riss93,jagau13} 
\begin{equation} \label{eq:cap2}
U = E - \eta \, dE/d\eta~.
\end{equation}
The optimal $\eta$ is then determined from $\text{min}[\eta \, dU/d\eta]$. For EOM-CCSD, $U$ can 
be computed analytically through $dE/d\eta = \text{Tr}[\gamma \, W]$ with $\gamma$ as one-particle 
density matrix.\cite{jagau13} As density matrices are not available for CAP-EOM-CCSD(T)(a)*, $U$ 
is computed via numerical differentiation of the energy at this level of theory. A formal inconsistency 
between CAP-EOM-CCSD and CAP-EOM-CCSD(T)(a)* arises because using a density matrix 
excluding orbital and amplitude response in the analytical evaluation of $U$ as proposed in Ref. 
\citenum{jagau13} corresponds to numerical differentiation of the energy with the CAP included only 
in the EOM-CC calculation. Some test calculations with a fully relaxed density matrix available from 
the CAP-EOM-CCSD analytic gradient code \cite{benda17} show, however, that this inconsistency 
is negligible in the computational practice. 

EOM-EA-CCSD*, EOM-EA-CCSD(fT), EOM-EA-CCSD(T)(a)* and their CAP-aug\-mented variants 
have been implemented into the \textsc{Q-Chem} program package.\cite{qchem} In addition, 
corresponding expressions for the EE and SF flavors of CAP-EOM-CC theory have been implemented. 
The implementation builds on the general implementation of CAP-EOM-CCSD in \textsc{Q-Chem} 
\cite{zuev14} and uses the libtensor library \cite{epifanovsky13} for high-performance tensor operations. 

\section{Application to Bound States} 
\label{sec:app1}
%In order to assess their numerical performance, 
In order to establish a method for the cost-effective treatment of radical anions beyond the 
EOM-EA-CCSD approximation, the various non-iterative triples corrections discussed in 
Section \ref{sec:th} as well as the EOM-EA-CCSD and EOM-EA-CC3 methods were employed 
to calculate vertical attachment energies of a test set of 22 molecules with bound EA states. 
Here, a rigorous assessment of the numerical performance is possible by comparing to 
EOM-EA-CCSDT results. While there are several test sets for the computation of electron 
affinities available from the literature \cite{curtiss98,boese01,curtiss05}, they typically 
comprise closed-shell and open-shell species and are therefore not well suited in the 
present context because EOM-EA-CC methods are preferably used starting from a 
closed-shell reference state.\cite{ccbook}

The test set put together for the present study comprises 13 neutral molecules (BH, LiH, ClF, 
Cl$_2$, CH$_2$, SiH$_2$, CHF, CHCl, CH$_3$Li, CH$_2$S, SO$_2$, S$_2$O, O$_3$) 
and 9 cations (CH$^+$, CF$^+$, NO$^+$, NO$_2^+$, NH$_2^+$, PH$_2^+$, CH$_3^+$, 
HCNH$^+$, CH$_2$NH$_2^+$). All these species have a closed-shell ground state and 
support a vertically bound electron-attached valence state at their equilibrium geometry. 
Remarkably, there are not many neutral species that share these two features as most small 
molecules with a closed-shell ground state either do not support a vertically bound radical 
anion at all or only a dipole-bound anion \cite{simons08}. In contrast, medium-size organic 
molecules feature bound valence radical anions more often, but are beyond the reach of 
EOM-EA-CCSDT and were for this reason not included here. 

All calculations were done at the equilibrium structures of the closed-shell reference states 
with the aug-cc-pV\textit{X}Z (\textit{X}=D, T, Q) basis sets \cite{kendall92,woon93} not including 
core electrons in the correlation treatment. Equilibrium structures were optimized at the 
CCSD/aug-cc-pCVTZ level of theory. EOM-EA-CCSD, -CCSD*, -CCSD(fT), and -CCSD(T)(a)* 
calculations were performed with a modified version of the \textsc{Q-Chem} program package 
\cite{qchem} while EOM-EA-CC3 and EOM-EA-CCSDT calculations were carried out with the 
implementations of the corresponding EOM-EE-CC methods in the \textsc{Cfour} program 
package \cite{cfour} including a continuum orbital in the basis set.\cite{stanton99} 

%as well as computational details
Table \ref{tab:bound} compiles mean and maximum deviations from EOM-EA-CCSDT obtained 
for the vertical attachment energies of the test set. Results for individual molecules are available 
in the supplementary material. Table \ref{tab:bound} illustrates that EOM-EA-CCSD* and 
EOM-EA-CCSD(fT) deviate roughly twice as much from EOM-EA-CCSDT as EOM-EA-CCSD 
and therefore cannot be recommended in the present context. A closer look at individual 
results reveals that both methods systematically overestimate the energy difference between 
the reference and the EA state since only the energy of the latter is corrected, whereas that 
of the former is still computed at the CCSD level. In contrast, EOM-EA-CCSD(T)(a)* reduces 
the mean absolute and maximum deviations from EOM-EA-CCSDT by 60-70\% and 80\%, 
respectively, compared to EOM-EA-CCSD and is of similar quality as or potentially slightly 
superior to EOM-EA-CC3, which treats triples excitations iteratively. Also, other than the 
EOM-EA-CCSD* and -CCSD(fT) corrections, the latter two methods do not deviate 
systematically in one direction from EOM-EA-CCSDT indicating a more balanced treatment 
of correlation in the reference and the target state. 

The present study demonstrates that the attachment energy of most molecules increases 
when going from EOM-EA-CCSD to EOM-EA-CCSDT. This is expected since inclusion of 
triples excitations should in general improve the description of the EOM state more than that 
of the reference state. There are, however, 6 species (O$_3^-$, SO$_2^-$, NO, NO$_2$, CF, 
HCNH) for which the opposite trend is observed. EOM-EA-CCSD(T)(a)* and EOM-EA-CC3 
capture this behavior for all cases but O$_3^-$ where all approximate triples methods do not 
improve upon EOM-EA-CCSD. This failure may be related to the sizable multiconfigurational 
character of the O$_3$ reference state \cite{hino06}; its description is significantly improved 
at the CCSDT level.  

Overall, Table \ref{tab:bound} suggests that EOM-EA-CCSD(T)(a)* offers a good balance 
between numerical accuracy and computational cost for vertical attachment energies of 
bound states beyond EOM-EA-CCSD. Since their electronic structure is similar, one can 
expect comparable performance for temporary radical anions of valence character, whereas 
dipole-bound \cite{simons08} and correlation-bound \cite{sommerfeld10} anions as well 
as dipole-stabilized resonances \cite{jagau15} may behave differently. 

Also, it is difficult to assess the performance of all EOM-EA-CC variants for doubly excited 
states because bound states of such character are exotic. In contrast, their temporary 
counterparts, that is, Feshbach resonances are of greater importance;\cite{jagau17} a 
respective pilot application is presented in Section \ref{sec:app3}.

%as their electronic structure is different. 

%However, there is one case that all approximate methods including EOM-EA-CC3 fail to 
%describe correctly: ozone. The attachment energy of this molecule decreases when going 
%from EOM-EA-CCSD to EOM-EA-CCSDT whereas all methods with approximate triples 
%excitations predict an increase. The decrease is counterintuitive since the correlation energy 
%should be larger for an ($N$+1) electron state than for an $N$ electron state, but can be 
%related to the well-established significant multireference character of neutral ozone. Notably, 
%there are 5 more species (SO$_2$, NO$^+$, NO$_2^+$, CF$^+$, HCNH$^+$) besides 
%O$_3$ in the test set for which EOM-EA-CCSDT yields a smaller attachment energy than 
%EOM-EA-CCSD, but this is reproduced by EOM-EA-CCSD(T)(a)* and EOM-EA-CC3.

% refer to Szalay paper about EE methods 
% molecular structures reported in supplement 
% why these systems? 
% --> size: need to be tractable by CCSDT
% --> el structure: needs to have closed-shell neutral ground state
% --> needs to be bound

%Still, as the focus of the present study is on temporary radical anions whose electronic structure 
%is similar to that of the molecules included in the test set. 
% Still, since the purpose of the test set is to establish a method for the treatment of residual electron correlation in temporary anions, 
%whose electronic structure is similar to that of the molecules included in the test set, the 
% justified because focus on resonances 

\section{Application to Shape Resonances}
\label{sec:app2}
To illustrate the impact of triples excitation on positions and widths of shape resonances, several 
temporary anions of $\pi^*$ (N$_2^-$, CO$^-$, C$_2$H$_2^-$, CH$_2$O$^-$, CO$_2^-$) and 
$\sigma^*$ (CH$_2$Cl$_2^-$) type were computed with CAP-EOM-EA-CCSD and -CCSD(T)(a)*. 
All calculations were carried out using the aug-cc-pV\textit{X}Z (\textit{X} = D, T, Q) basis sets 
augmented by additional even-tempered diffuse functions.\cite{zuev14} These extra basis functions 
were placed in the center of the molecule (denoted (C) in Table \ref{tab:res}) in the cases of N$_2^-$, 
CO$^-$, C$_2$H$_2^-$, and CH$_2$O$^-$ where the electron is attached to an isolated double 
bond and at all heavy atoms for CO$_2^-$ and CH$_2$Cl$_2^-$ (denoted (A) in Table \ref{tab:res}). 
Core electrons were frozen in all calculations. Molecular structures and further computational 
details such as CAP onsets and optimal CAP strengths are documented in the supplementary 
material. 

The results in Table \ref{tab:res} show that CAP-EOM-EA-CCSD(T)(a)* in general yields lower 
resonance positions than CAP-EOM-EA-CCSD. This agrees well with the change in the same 
direction when going from CAP-HF to CAP-EOM-EA-CCSD \cite{jagau17} and also with the opposite 
trend observed for the attachment energies of most bound EA states (cf. Section \ref{sec:app1}). 
The extent of the lowering is similar for uncorrected and first-order corrected values, but varies 
between different resonances. For N$_2^-$, CH$_2$O$^-$, and CH$_2$Cl$_2^-$ it amounts 
to about 0.05 eV and is thus comparable to the size of the first-order correction for the CAP (cf. 
Eq. \eqref{eq:cap2}). For CO$^-$, C$_2$H$_2^-$, and CO$_2^-$ on the other hand, the difference 
is at the order of 0.01 eV and therefore negligible for most practical purposes. 

Table \ref{tab:res} also demonstrates that going from CAP-EOM-EA-CCSD to 
CAP-EOM-EA-CCSD(T)(a)* lowers the widths of all six resonances studied here. This is 
somewhat counterintuitive since CAP-EOM-EA-CCSD yields significantly larger resonance 
widths than CAP-HF \cite{jagau17} and one might infer that a more complete description of 
electron correlation further increases the resonance width. As for resonance positions, the 
lowering is similar for uncorrected and first-order corrected values, but not for the different 
species. Notably, the difference between the two methods is largest for CO$^-$ (about 0.05 
eV) whose position changes only to a negligible extent. For N$_2^-$ and CH$_2$O$^-$ the 
difference is also not entirely insignificant, whereas almost identical resonance widths are 
obtained in the cases of C$_2$H$_2^-$, CO$_2^-$, and CH$_2$Cl$_2^-$. 

\begin{figure}[htbp]
\includegraphics[scale=0.895]{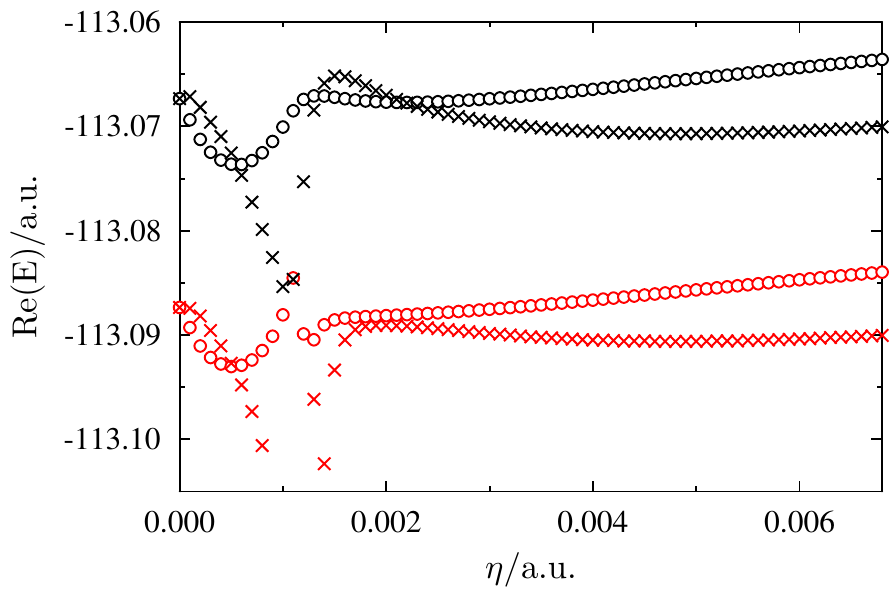} 
\includegraphics[scale=0.895]{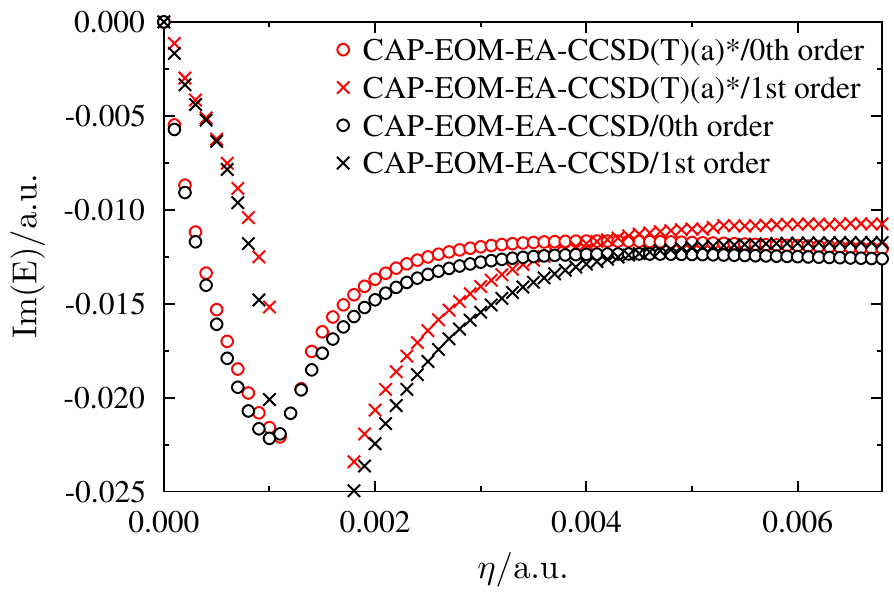}
\caption{Real (left) and imaginary (right) energy of the $^2\Pi$ shape resonance of CO$^-$ as a function 
of CAP strength computed at the CAP-EOM-EA-CCSD and -CCSD(T)(a)* levels of theory using the 
aug-cc-pVTZ+6s6p6d(C) basis set.}
\label{fig:shape} \end{figure}

Since triples excitations do not change the positions and widths of shape resonance substantially, 
one may wonder if it possible to save computer time by determining the optimal CAP strength at 
the EOM-EA-CCSD level of theory followed by a single CAP-EOM-EA-CCSD(T)(a)* calculation. 
This is appears to be a valid approach given the very similar optimal CAP strengths obtained with 
CAP-EOM-EA-CCSD and -CCSD(T)(a)* (see supplementary material) and is further substantiated 
by the fact that both methods produce similar $\eta$-trajectories. This is documented in Figure 
\ref{fig:shape}, which displays $\eta$-trajectories for CO$^-$ as a representative example. 

Overall, one can conclude from Table \ref{tab:res} that triples excitations make a minor but in 
many cases non-negligible impact on positions and widths of shape resonances. In particular, 
resonance positions change to a similar extent as attachment energies of bound EA states but 
in the opposite direction. These numerical findings confirm the expectation that resonances with 
dominant single-attachment character should be described accurately within the EOM-EA-CCSD 
approximation. 

Further trends observed in Table \ref{tab:res}, i.e., with respect to basis-set size and the first-order 
correction are in line with previous findings and need not be discussed in detail. It is, however, 
worth noting that there are considerable discrepancies between the best theoretical estimates 
in Table \ref{tab:res} and the corresponding experimental values. For example, the 
CAP-EOM-EA-CCSD(T)(a)*/aug-cc-pVQZ+6s6p6d(C) results for N$_2^-$ including the first-order 
correction are still 0.1 eV and 0.14 eV away from the fixed-nuclei estimate obtained through 
a fit to experimental data (E = 2.32 eV, $\Gamma$ = 0.41 eV).\cite{berman83} For CO$^-$, 
the difference is even larger: The CAP-EOM-EA-CCSD(T)(a)*/aug-cc-pVQZ+6s6p6d(C) values 
are E = 1.915 eV and $\Gamma$ = 0.657 eV whereas a fit to experimental data yielded E = 
1.50 eV and $\Gamma$ = 0.75--0.80 eV.\cite{zubek77,zubek79} The situation is similar for the 
other resonances, but the comparison is more difficult for polyatomic species where fixed-nuclei 
estimates are harder to deduce from experimental data. From Table \ref{tab:res} it is clear that 
higher-order electron correlation does not provide a sufficient explanation for these discrepancies. 
Rather, incompleteness of the one-electron basis set may explain part of the difference as 
resonance positions and widths obtained with the modified aug-cc-pVQZ basis set are likely 
not yet converged. Also, results obtained at the same level of correlation treatment but using 
other techniques than CAPs, for example, complex basis functions \cite{white17} or the 
stabilization method \cite{falcetta14} are significantly different in some cases.
% Although a comprehensive discussion of the origin of these discrepancies is beyond the present work, 

As a final remark to Table \ref{tab:res}, an ambiguity in the determination of the optimal CAP 
strength for N$_2^-$/aug-cc-pVTZ+6s6p6d(C) should be addressed. Two minima in both $\eta \, 
dE/d\eta$ and $\eta \, dU/d\eta$ are found for this particular system and basis set. The values 
included in Table \ref{tab:res} conform to results obtained with the other basis sets, but the 
other set of values, i.e., the ones included as a footnote to Table \ref{tab:res}, are closer to the 
estimate inferred from experimental data and were also reported in Ref. \citenum{jagau16}. 
A rigorous solution to this ambiguity would demand the use of another basis set, but for the 
present purpose, i.e., investigating the impact of triples excitations, this did not seem to be 
necessary.

\begin{figure}[htbp]
\includegraphics[scale=1.2]{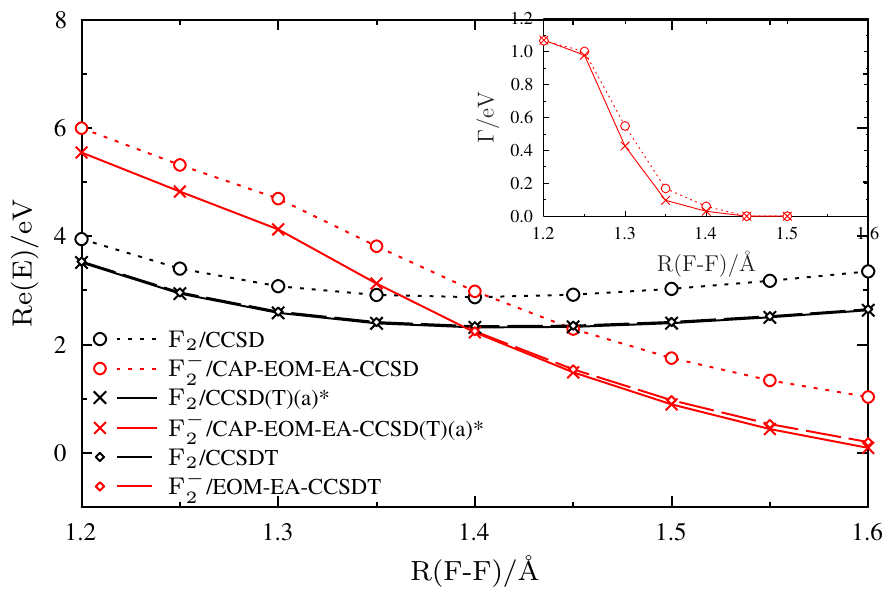} 
\caption{Potential energy curves of F$_2^-$ computed at the CAP-EOM-EA-CCSD and -CCSD(T)(a)* 
levels of theory using the aug-cc-pVTZ+6s6p(A) basis set including first-order correction for the CAP. 
The corresponding curves of neutral F$_2$ computed with CCSD and CCSD(T)(a) are also shown.}
\label{fig:f2pec} \end{figure}

As a further application, the potential energy curve of the $\sigma^*$ resonance of F$_2^-$ and 
its conversion into a bound anion were studied at the CAP-EOM-EA-CCSD and -CCSD(T)(a)* 
levels of theory. The bound part of the potential energy curve was additionally computed with 
EOM-EA-CCSDT. This is documented in Figure \ref{fig:f2pec}. While triples excitations do not 
change the description of F$_2^-$ qualitatively, this application illustrates some further aspects 
of the CAP-EOM-CCSD(T)(a)* method. First, the conversion from temporary to bound anion is 
not described consistently. At R(FF) = 1.4 \AA, the anion is computed to be lower in energy than 
the neutral molecule, but one still obtains a finite resonance width of about 0.03 eV. If the 
Schr\"odinger equation was solved exactly, the resonance width would become zero at the same 
bond distance where the potential curves of F$_2^-$ and F$_2$ cross. This feature is preserved 
in CAP-EOM-CCSD where a temporary anion and its parent state are obtained as eigenfunctions 
of the same Hamiltonian \cite{jagau14}, but not in CAP-EOM-CCSD(T)(a)* due to the correction 
of the target-state energy according to Eq. \eqref{eqdelta}. 

Moreover, the bound part of the F$_2^-$ potential energy curve in Figure \ref{fig:f2pec} demonstrates 
that EOM-EA-CCSD(T)(a)* reproduces EOM-EA-CCSDT total energies well, but it is also seen that 
the deviation grows with increasing bond length. In fact, at R(FF) = 1.6 \AA~EOM-EA-CCSD(T)(a)* 
overestimates the attachment energy already by about 0.09 eV relative to EOM-EA-CCSDT, whereas 
the methods agree up to 0.01 eV at R(FF) = 1.4 \AA. This indicates that the increasing multiconfigurational 
character of neutral F$_2$ at stretched bond lengths is not fully captured within the CCSD(T)(a) 
approximation. 

\section{Application to Feshbach Resonances}
\label{sec:app3}
As an example of a Feshbach resonance, the $^2\Sigma^+$ state of CO$^-$ at around 10 eV 
was investigated. This resonance arises through attachment of a $3s \, \sigma_g$ electron to the 
$b\; ^3\Sigma^+$ or the $B\; ^1\Sigma^+$ Rydberg state of neutral CO, but lies energetically below 
both these parent states so that it can decay only through a two-electron process. In a CAP-EOM-CC 
treatment based on the ground state of neutral CO as reference, the resonance wave function is 
dominated by a double excitation and it can be anticipated that CAP-EOM-EA-CCSD places the 
resonance significantly too high in energy, i.e., above its parent states turning it into a core-excited 
shape resonance. Whether CAP-EOM-EA-CCSD(T)(a)* produces the correct energetic order is less 
clear and investigated here. 

For a number of reasons, the $^2\Sigma^+$ resonance of CO$^-$ represents a good test case 
for exploratory calculations: Firstly, reliable experimental data about its position (10.04 eV) and 
width (0.04 eV) have been reported \cite{schulz73} and, secondly, its wave function is largely 
dominated by the aforementioned single configuration. This is in contrast to Feshbach resonances 
in medium-sized molecules where configuration mixing can entail appreciable single-attachment 
character. However, describing electron attachment to Rydberg states requires some changes 
to the computational protocol established for valence shape resonances:\cite{zuev14} The 
d-aug-cc-pV\textit{X}Z basis sets \cite{woon94} have to be used instead of aug-cc-pV\textit{X}Z 
to obtain converged energies for the parent Rydberg states and moreover, the CAP onset has 
to be chosen much larger than the spatial extent of the neutral ground state and also that of the 
parent states in order to limit the perturbation of the latter states to an acceptable level. 
%Details are documented in the supplementary material. 
%no new diff basis functions

\begin{figure}[htbp]
\includegraphics[scale=0.81]{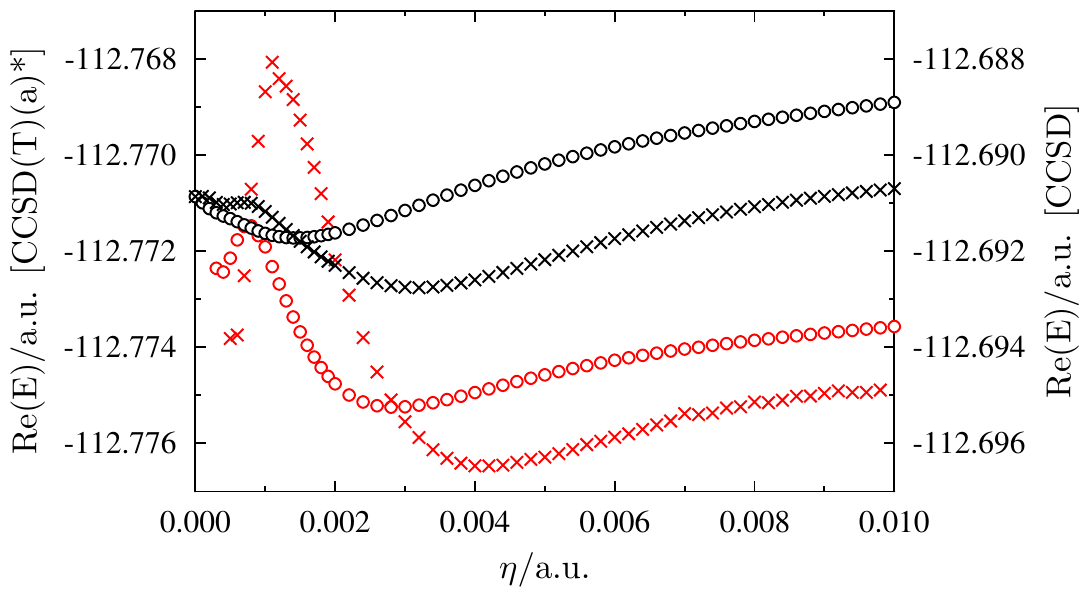}
\includegraphics[scale=0.81]{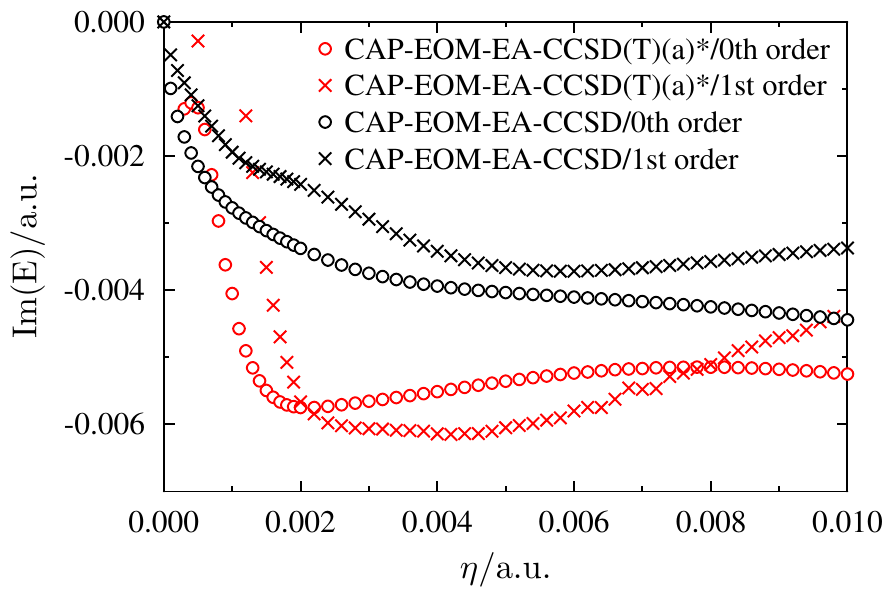}
\caption{Real (left) and imaginary (right) energy of the $^2\Sigma^+$ Feshbach resonance of CO$^-$ 
as a function of CAP strength computed at the CAP-EOM-EA-CCSD and -CCSD(T)(a)* levels of theory 
using the d-aug-cc-pVTZ+6s6p(C) basis set. Note the different energy scales for the two methods in 
the left panel.}
\label{fig:fesh} \end{figure}

Figure \ref{fig:fesh} displays $\eta$-trajectories for the $^2\Sigma^+$ resonance of CO$^-$ 
obtained with CAP-EOM-EA-CCSD and CAP-EOM-EA-CCSD(T)(a)* using the d-aug-cc-pVTZ+6s6p(C) 
basis set. The differences between the curves illustrate that including triples excitations changes 
the description of this system qualitatively. At the CAP-EOM-EA-CCSD level, the resonance 
position and width are obtained as 12.38 eV and 0.17 eV. Since the $b\; ^3\Sigma^+$ parent state 
is placed at 10.54 eV by EOM-EE-CCSD, the decay via a one-electron process is possible, which 
explains the unphysically large width. At the CAP-EOM-EA-CCSD(T)(a)* level, the resonance 
position drops to 10.58 eV and the deviation from the experimental value decreases from over 
2 eV to 0.54 eV. However, despite this improvement, the description is still qualitatively wrong 
because the resonance is above its parent state, which is placed at 10.32 eV by EOM-EE-CCSD(T)(a)*. 
Consequently, the resonance width is again unphysically large (0.33 eV).

This failure illustrates that CAP-EOM-EA-CCSD(T)(a)* results cannot be trusted when triples 
excitations induce a qualitative change in the wave function. It is, however, not surprising because 
the method is explicitly designed for the treatment of states with single excitation/attachment 
character.\cite{matthews16} The perturbative analysis of $\Delta E$ (Eqs. \eqref{eqdelta}--\eqref{eqft2}) 
assumes that $R_1$ contributes at zeroth order but $R_2$ only at first order, which is not the case 
for Feshbach resonances. 

A consistent solution to this problem is likely afforded by the full inclusion of triples excitations, 
i.e., by CAP-EOM-EA-CCSDT. This entails, however, considerably increased computational 
cost and is beyond the scope of the present work. Alternatively, the parent triplet Rydberg state 
could be used as reference state in an EOM-CCSD treatment similar to what has been done 
for resonances of DNA bases in the context of the stabilization method.\cite{fennimore16} Such 
an approach would provide a more balanced treatment as the Feshbach resonance is of 
single-attachment character and the ground state of the neutral molecule is described through 
a spin-flipping single excitation, but it is state-specific in the sense that different resonances 
of the same system are described based on different reference states. 
% alternative: CAS-SCF or SF-CCSD
% both have their own drawbacks

% first time triples for resonances, no rigorous comparison possible because nothing for resonances beyond CCSD
% impact of T on widths unclear
% CO and N2 isoelectronic but impact on E of N2 more significant

\section{Concluding Remarks}
\label{sec:con}

In this article, electronic resonances have been studied for the first time beyond the EOM-CCSD 
approximation. To avoid the high cost of full EOM-CCSDT, triples excitations are taken into 
account in a non-iterative manner. Since none of the methods proposed for doing so has been 
applied previously to the EA variant of EOM-CC, benchmark calculations for several bound 
electron-attached states were carried out first with EOM-EA-CCSDT as reference point. 

EOM-EA-CCSD* and EOM-EA-CCSD(fT), which correct only target states, do not yield accurate 
attachment energies, although significant improvements over EOM-CCSD have been reported 
for vertical ionization potentials \cite{manohar09} and energy differences between EOM-SF states 
\cite{manohar08}. In contrast, the EOM-EA-CCSD(T)(a)* method, in which the CCSD reference 
state is corrected for the effect of triples excitations before constructing the similarity-transformed 
Hamiltonian, reliably improves upon EOM-EA-CCSD attachment energies and is comparable in 
accuracy to EOM-EA-CC3.

Selected applications of EOM-EA-CCSD(T)(a)* augmented by a CAP to temporary anions with 
shape-resonance character illustrate that higher-order electron correlation uniformly lowers their 
positions and widths somewhat, in the case of F$_2^-$ even by more than 0.1 eV. The present 
work thus confirms that CAP-EOM-EA-CCSD describes shape resonances rather accurately, 
but also demonstrates that the effect of higher excitations is not insignificant in many cases but 
of similar magnitude as the first-order correction for the CAP perturbation. 
%and changes induced by variations of the CAP onset parameters.\cite{zuev14} 

While CAP-EOM-EA-CCSD(T)(a)* allows to reliably assess higher-order correlation effects in 
shape resonances at acceptable computational cost (non-iterative $\mathcal{O}(N^7)$), results 
for Feshbach resonances are not satisfactory. As these states are doubly excited with respect 
to the CC reference state, inclusion of triples excitations changes their wave function qualitatively, 
which is not captured by CAP-EOM-EA-CCSD(T)(a)*: Feshbach resonances are substantially 
lowered in energy compared to CAP-EOM-EA-CCSD, but still appear above their parent states 
and thus described incorrectly as core-excited shape resonances. This demonstrates that a more 
complete treatment of triples excitations is required to achieve a consistent description of Feshbach 
resonances and their parent states in the context of EOM-CC. 

In sum, CAP-free and CAP-augmented EOM-EA-CCSD(T)(a)* could be established as reliable 
methods for bound and temporary anions of single-attachment character. Exemplary applications 
illustrate that triples excitations lower resonance positions and widths and confirm the validity 
of the CAP-EOM-EA-CCSD approach.
% to assess the impact of residual electron correlation beyond the singles and doubles approximation on

\section*{Supplementary Material}
See supplementary material for computational details such as molecular structures, and CAP 
parameters as well as results corresponding to Table \ref{tab:bound} and Figure \ref{fig:f2pec}. 

\section*{Acknowledgments}
The author thanks Devin Matthews for help in verifying the implementation of the EOM-CCSD(T)(a)* 
method. Financial support by the Fonds der Chemischen Industrie through a Liebig fellowship is 
gratefully acknowledged. 
%The author thanks Professor Christian Ochsenfeld for hospitality and support at LMU Munich. 

\clearpage

\section{Tables}

\begin{table}[h] \centering
\caption{Absolute deviations, mean absolute deviations, and maximum deviations in eV of vertical 
attachment energies of a test set of 22 molecules computed at various levels of EOM-EA-CC theory 
from the corresponding EOM-EA-CCSDT values.}
\vspace{0.2cm} 
\renewcommand{\arraystretch}{1.3}
\setlength{\tabcolsep}{5pt}
\begin{tabular}{cccccc} \hline \hline
 & \multicolumn{5}{c}{EOM-EA-} \\
basis & CCSD & CCSD* & CCSD(fT) & CCSD(T)(a)* & CC3 \\ \hline
 & \multicolumn{5}{c}{mean deviation} \\
 aug-cc-pVDZ & 0.021 & -0.103 & -0.078 & -0.006 & -0.003 \\
 aug-cc-pVTZ & 0.014 & -0.118 & -0.088 & -0.008 & -0.002 \\
 aug-cc-pVQZ & 0.011 & -0.123 & -0.092 & -0.011 & -0.002 \\ \hline
 & \multicolumn{5}{c}{mean absolute deviation} \\
aug-cc-pVDZ & 0.036 & 0.104 & 0.082 & 0.013 & 0.020 \\
aug-cc-pVTZ & 0.041 & 0.118 & 0.091 & 0.014 & 0.019 \\
aug-cc-pVQZ & 0.042 & 0.123 & 0.095 & 0.016 & 0.018 \\ \hline
 & \multicolumn{5}{c}{maximum deviation$^a$} \\
aug-cc-pVDZ & 0.114 & 0.286 & 0.271 & 0.022 & 0.048 \\
aug-cc-pVTZ & 0.147 & 0.320 & 0.297 & 0.025 & 0.063 \\ 
aug-cc-pVQZ & 0.157 & 0.329 & 0.304 & 0.030 & 0.067 \\ \hline \hline
\end{tabular} 
\footnotetext{excluding O$_3$}
\label{tab:bound} \end{table}

\begin{sidewaystable} \centering
\caption{Resonance positions and widths in eV of several valence shape resonances computed with  
CAP-EOM-EA-CCSD and CAP-EOM-EA-CCSD(T)(a)* using different basis sets.}
\vspace{0.1cm} \begin{small}
\renewcommand{\arraystretch}{1.3}
\setlength{\tabcolsep}{4.5pt}
\begin{tabular}{ll|cccc|cccc} \hline\hline
 &  & \multicolumn{4}{c|}{Resonance positions} & \multicolumn{4}{c}{Resonance widths} \\ \hline
 &  & \multicolumn{2}{c}{without correction for CAP} & \multicolumn{2}{c|}{including first-order} & 
 \multicolumn{2}{c}{without correction for CAP} & \multicolumn{2}{c}{including first-order} \\
 &  &  &  & \multicolumn{2}{c|}{correction for CAP} &  &  & \multicolumn{2}{c}{correction for CAP} \\ \hline
 &  & \multicolumn{2}{c}{CAP-EOM-EA-} & \multicolumn{2}{c|}{CAP-EOM-EA-} & 
 \multicolumn{2}{c}{CAP-EOM-EA-} & \multicolumn{2}{c}{CAP-EOM-EA-}\\
Molecule & Basis & CCSD & CCSD(T)(a)* & CCSD & CCSD(T)(a)* & CCSD & CCSD(T)(a)* & CCSD & CCSD(T)(a)* \\ \hline\hline
N$_2^-$ & aug-cc-pVDZ+6s6p6d(C) & 2.782 & 2.718 & 2.735 & 2.674 & 0.397 & 0.382 & 0.267 & 0.243 \\
N$_2^-$ & aug-cc-pVTZ+6s6p6d(C)$^a$ & 2.608 & 2.556 & 2.567 & 2.517 & 0.373 & 0.359 & 0.283 & 0.259 \\
N$_2^-$ & aug-cc-pVQZ+6s6p6d(C) & 2.510 & 2.467 & 2.461 & 2.412 & 0.359 & 0.340 & 0.300 & 0.273 \\ \hline
CO$^-$ & aug-cc-pVDZ+6s6p6d(C) & 2.292 & 2.249 & 2.161 & 2.155 & 0.728 & 0.697 & 0.695 & 0.635 \\
CO$^-$ & aug-cc-pVTZ+6s6p6d(C) & 2.114 & 2.105 & 2.021 & 2.018 & 0.675 & 0.631 & 0.646 & 0.591 \\
CO$^-$ & aug-cc-pVQZ+6s6p6d(C) & 2.023 & 2.021 & 1.903 & 1.915 & 0.710 & 0.660 & 0.715 & 0.657 \\ \hline
C$_2$H$_2^-$ & aug-cc-pVDZ+6s6p6d(C) & 2.871 & 2.860 & 2.792 & 2.786 & 0.878 & 0.870 & 0.613 & 0.599 \\
C$_2$H$_2^-$ & aug-cc-pVTZ+6s6p6d(C) & 2.689 & 2.683 & 2.507 & 2.505 & 0.968 & 0.947 & 0.863 & 0.847 \\
C$_2$H$_2^-$ & aug-cc-pVQZ+6s6p6d(C) & 2.452 & 2.455 & 2.202 & 2.207 & 1.135 & 1.118 & 1.009 & 1.009 \\ \hline
CH$_2$O$^-$ & aug-cc-pVTZ+6s6p6d(C) & 1.256 & 1.206 & 1.208 & 1.163 & 0.328 & 0.292 & 0.279 & 0.239 \\ \hline
CO$_2^-$ & aug-cc-pVTZ+6s6p6d(A) & 4.008 & 3.997 & 3.983 & 3.970 & 0.214 & 0.216 & 0.190 & 0.182 \\ \hline
CH$_2$Cl$_2^-$ & aug-cc-pVTZ+6s6p(A) & 1.941 & 1.902 & 2.019 & 1.975 & 0.732 & 0.726 & 0.565 & 0.572 \\ \hline \hline
\end{tabular} \end{small}
\footnotetext{A second set of values can be obtained due to an ambiguity in the determination of the optimal CAP 
strength for N$_2^-$/aug-cc-pVTZ+6s6p6d(C). These values are 2.516, 2.469, 2.461, 2.418 eV for the resonance 
position and 0.417, 0.385, 0.433, 0.385 eV for the resonance width.}
\label{tab:res} \end{sidewaystable}

\clearpage

\section{Figures}

%\clearpage 

\end{document}